\documentclass[aps,prb,twocolumn,showpacs]{revtex4}
\usepackage{graphicx} 
\begin{document}
\title
{\bf When is the two-level approximation untenable in issues of
Decoherence?}
\author{Kerim Savran$^{\bf (1)}$, T. Hakio\u glu$^{\bf (1)}$ and
E. Me\d se$^{\bf (2)}$}
\address{
${\bf (1)}$ {Department of Physics, Bilkent University, Bilkent,
06800 Ankara, Turkey}\break ${\bf (2)}$ {Department of Physics,
Faculty of Science and Art, Dicle University, Diyarbak\i r,
Turkey}}

\begin{abstract}

We examine the conditions in favor and necessity of a realistic
multileveled description of a decohering quantum system. Under
these conditions approximate techniques to simplify a multileveled
system by its first two levels is unreliable and a realistic
multilevel description in the formulation of decoherence is
unavoidable.

In this regard, our first crucial observation is that, the
validity of the two level approximation of a multileveled system
is not controlled purely by {\it sufficiently low temperatures}.  
We demonstrate using three different environmental spectral models 
that the type of system-environment coupling and  
the environmental spectrum have a dominant role over the  
 temperature. Particularly, zero temperature   
quantum fluctuations induced by the Caldeira-Leggett type
linear coordinate coupling  can be influential
in a wide energy range in the systems allowed transitions. The
second crucial observation against the validity of the two level
approximation is that the decoherence times being among the system's
short time scales are found to be dominated not by the resonant
but {\it non-resonant} processes. We demonstrate this in three stages. 
Firstly, our zero temperature numerical calculations 
reveal that, the calculated decoherence
rates including relaxation, dephasing and leakage phenomena show,  
a linear dependence on the spectral area for all spectral models used, 
independent from the 
spectral shape within a large environmental spectral range compared 
to the quantum system's energies. Secondly, within the same range, the 
decoherence times only have a marginal dependence 
on the translations of the entire frequency spectrum. Finally, the
same decoherence rates show strong dependence on the number
of coupled levels by the system-environment coupling.
\end{abstract}


\maketitle 
\section{Introduction}
The nature of the environmental coupling and the properties of the 
environmental spectrum are essential factors in the determination 
of the decoherence
properties of multileveled quantum systems (MLS). Most of the
realistic MLS have Caldeira-Leggett type linear    
coupling\cite{CL,LG} of a macroscopic system coordinate to an external
environmental noise field. In this publication, the MLS is
identified by a large number of energy levels coupled by
non-negligible matrix elements induced by an environmental
perturbation. The first two levels comprise the non ideal qubit
space which we consider to be energetically well separated from
the higher levels. The analysis of the decoherence properties of such 
MLS has 
recently appeared in a few publications\cite{TS1,DiVinc}.

Normally a physical environmental spectrum (the so called noise)
has a large range of frequencies with different characteristic behavior 
in different spectral regions. Moreover, the spectral noise range  
 usually extends well beyond the energy scales of the quantum
system that it interferes with. In most of the environmental
models, weak noise signals and a realistic linear coupling suffices
to model the system-noise interference. On the other hand, it is
customary to consider the decohering quantum system as a two level
approximated version of a generic MLS. The well studied
spin-boson\cite{SB1,SB2} or central spin\cite{SBCS} models are such
examples arising from the assumption that for sufficiently low
temperatures $T \ll \Delta E$ a MLS is well approximated by its first 
two levels\cite{SB1} (where $\Delta E$ is a characteristic energy scale 
separating the first two levels from the higher excited states of the 
quantum system). This two level approximation (2LA) has in its
essence three assumptions: a) that the incoherent transitions caused
by the environment in the system are generated by the resonant
processes: Implicit in the 2LA is the belief that the spectrum must 
have non-negligible couplings at the {\it right transition frequencies} at 
which the system makes transitions to higher levels. 
 b) at zero or sufficiently low temperatures there are no
available environmental states to couple with the system: This assumption 
and the notion of the {\it right transition frequency} basically help to 
neglect all parts of the spectrum $\Delta E \le \omega$ because of the  
religious belief that at sufficiently low temperatures the relevant part 
of the spectrum is the low energy sector $\omega \le T\ll\Delta E$ of which 
coupling is suppressed by thermal considerations. 
c) negligible 
leakage of the wave-function to higher levels: In fact these effects have 
not been seriously questioned until recently. Contrary arguments to 
the negligible leakage can be found for instance in the recent 
publication Ref.[\cite{DiVinc}]. 

This manuscript presents counter arguments to these three fundamental 
assumptions of the 2LA mentioned above by considering a
MLS interacting with a bosonic environmental bath at zero
temperature through a Caldeira-Leggett type linear coordinate coupling. 
By considering a spectral   
noise range extending well beyond the thermal and the system energy scales,  
and adopting a model MLS for which $T \ll \Delta E$ is well satisfied we 
demonstrate that the three basic assumptions of the 2LA are neither 
necessary nor sufficient for its validity.  

The method that we employ is the direct numerical solution of the
master equation for the multileveled quantum systems' reduced density matrix
(RDM) in the short-time Born-Oppenheimer approximation. That the calculated
decoherence rates are largely independent from where the energy
levels are, is demonstrated. This observation of the independence of 
the decoherence rates from the location of the resonant transition
energies provides the first clue of the strong influence of the
non-resonant processes. The main results are two fold: i)
Decoherence is dominated by the non-resonant processes at zero temperature  
implying that
decoherence times are finite and saturate near zero temperature: This 
observation basically invalidates the assumption (a) above. 
ii) In a realistically multileveled system, and following (i),     
non-resonant processes  
induced in the entire environmental spectral range additionally 
 cause wave-function leakage from the qubit subspace to the
system's higher levels. Our numerically calculated leakage (L)
times are in the same order of magnitude as the
relaxation/dephasing (RD) times. 
Therefore the leakage, being a characteristic MLS effect, renders
itself to be non-negligible at short times. The observation of 
strong leakage at short times invalidates the conventional belief that 
the affect of 
multilevels in decoherence is a mere renormalizion of the effective two  
level system parameters.  

The zero temperature quantum fluctuations of the model system in 
(\ref{hamilt.0}) are caused by the
system-environment coupling in which the number of excitations in
the environmental modes is unconserved by the interaction.
 Decoherence is caused by the participation of large
number of these modes in the interaction. 
Similar zero temperature decoherence mechanisms
have been recently verified for the mesoscopic persistent
current rings experimentally\cite{PCdecoh1} and studied
theoretically\cite{PCdecoh2}. In particular, the    
saturation observed in the electron dephasing time in disordered 
conductors\cite{LinGiordano} has been argued in favour of the 
zero temperature quantum fluctuations\cite{MJW}. 

In Section II.A the observation (i) is demonstrated starting from a  
finite level system-environment model. Most of the results are confined 
to three levels. Section II.B is devoted to the explicit multilevel  
effects which demonstrate the observation (ii). In section II.C the 
influence of the non-resonant processes and the multilevels on the noise 
vacuum fluctuations are examined.

\section{A multilevel system-environment model} 
At the first step we show the strong influence of the
non-resonant transitions by finding that the decoherence times are
basically independent from the shape and the location of the spectrum. 
The calculated relaxation, dephasing and leakage (RDL) rates are
linearly dependent on the spectral area in a large range of
environmental frequencies. Therefore the participation of the
system's higher transition levels is unavoidable.
In the second step,
we explicitly calculate the RDL rates as function of the number of
levels in MLS and demonstrate that the higher transition levels
can strongly affect the rates.

Although some results are presented for $M$ leveled MLS, most of 
our conclusions are based on the solution of the
three level system-environment Hamiltonian (i.e. $M=3$) 
\begin{equation}
\begin{array}{rl}
{\cal H}=&\sum_{n=1}^{3}E_n\,\vert n\rangle\langle n \vert+ 
\int_{\omega}d\omega\, \omega \,(a_\omega^\dagger a_\omega+
a_\omega a_\omega^\dagger) \\ 
&\\
&+\sum_{n,r=1}^{3}\varphi_{n\,r}\vert n\rangle\langle r\vert \int
d\omega\, \eta_\omega (a_\omega+a_\omega^\dagger)
\label{hamilt.0}
\end{array}
\end{equation}
Here $E_1=E_2=0.5$ and $E_3=2.5$ are fixed system eigenenergies in some
absolute (and irrelevant) energy scale and $\vert r\rangle$ indicates
the $r$'th system eigenstate. We consider the couplings
$\varphi_{nr}$ to be real-symmetric and
$\varphi_{12}=\varphi_{23}=0.1$, and $\varphi_{13}=0$
 in the same absolute energy unit. The frequency dependent
part of the system-environment coupling is separately denoted by
$\eta_\omega$. All other energy (time) scales -including the
spectral energy scales
 are dimensionless in the same absolute energy (time)
scale. The environment is characterized by a spectral function
$I(\omega)=\eta^2_\omega\,(2n_{\omega}+1)$ where $n_{\omega}$ is
the bosonic thermal distribution. Since our results are confined
to zero temperature the bare distribution $n_{\omega}=0$, and   
therefore $I(\omega)=\eta^2_\omega$
for which we use two spectral models below. For obtaining results
in regard of the pure two level system we simply reset the couplings to
the third level in (\ref{hamilt.0}) to zero (i.e.
$\varphi_{13}=\varphi_{23}=0$). From the 2LA point of view, we have  
a quantum system (\ref{hamilt.0}) which is supposed to be a good candidate 
for a two level system (Since $\Delta E=E_3-E_2 \ne 0$ and the 
environmental 
temperature is zero; i.e. $T \ll \Delta E$ is well satisfied). However, 
as we show below, the model (\ref{hamilt.0}) breaks all three aspects  
of the 2LA.  

The machinery that we use is the master equation formalism in the 
interaction picture. The
quantum system is assumed to be initially prepared in the qubit subspace
$\vert \psi(0)\rangle= a \vert 1\rangle+ b \vert 2\rangle$ where we 
consider $a=\sqrt{0.1}, b=e^{i\pi/2}\,\sqrt{0.9}$.
For the short time dynamics it is sufficient to use the
Born-Oppenheimer approximation\cite{BornOppenheimer}
in the time evolution of the density matrix
 $\hat{\tilde{\rho}}(t)=\hat{\tilde{\rho}}^{(S)}(t)\otimes
\hat{\tilde{\rho}}_{e}(0)$. Here $\tilde{}$ denotes the
interaction picture, $\rho^{(S)}$ denotes the system's reduced
density matrix (RDM) and $\rho_e$ is the density matrix of the
pure environment characterized by the spectral function $I(\omega)$.  
The master equation for the system's RDM is\cite{TS1}  
\begin{equation}
\frac{d}{dt}{\tilde{\rho}}_{nm}^{(S)}(t)=-\int_{0}^{t}\,dt^\prime\,
\sum_{r,s}\,K_{rs}^{nm}(t,t^\prime)\,
\tilde{\rho}_{rs}^{(S)}(t^\prime)
\label{densmatr.2}
\end{equation}
where
\begin{equation}
\begin{array}{lrlr}
&&K_{rs}^{nm}(t,t^\prime) \\
&&=\Bigl\{{\cal F}(t-t^\prime)
[({\tilde{\varphi}}_t{\tilde{\varphi}}_{t^\prime})_{n\,r}
\delta_{s,m}- ({\tilde{\varphi}}_{t^\prime})_{n\,r}
({\tilde{\varphi}}_{t})_{s\,m}] \\
&&+{\cal F}^*(t-t^\prime)
[({\tilde{\varphi}}_{t^\prime}{\tilde{\varphi}}_{t})_{m\,s}
\delta_{r,n}-({\tilde{\varphi}}_{t})_{n\,r}
({\tilde{\varphi}}_{t^\prime})_{s\,m}]\Bigr\}
\end{array}
\label{densmatr.3}
\end{equation}
is the non-Markovian system-noise kernel. Here 
$(\tilde{\varphi}_t)_{n\,r}=(\tilde{\varphi}_0)_{n\,r}\,e^{-i(E_n-E_r)t}$. 
The noise correlator
${\cal F}(\tau)= \Theta(\tau)\,\int_{-\infty}^{\infty} d\omega
e^{i\omega \tau} I(\omega)$ is calculated for three different types
of spectral functions: a) a rectangular spectrum
\begin{equation}
I_{rec}(\omega)=\frac{A}{2\epsilon}\Theta(\epsilon^2-(\omega-\omega_0)^2)
\label{noisecorrelator.1}
\end{equation}
and b) a Lorentzian spectrum
\begin{equation}
I_{Lor}(\omega)=
\frac{A}{\pi}\frac{\epsilon}{(\omega-\omega_0)^2+\epsilon^2}
\label{noisecorrelator.2}
\end{equation}
and the power-Gaussian spectrum 
\begin{equation}
I(\omega)={\cal A}\,(\frac{\omega}{\Lambda})^{1+\nu}\,e^{-\omega^2/4\Lambda^2}
\label{powGauss}
\end{equation}
The common parameters in the first two spectra are the location of the 
center of
the spectra $\omega_0$ and the spectral width $\epsilon$. The
spectral area in both cases is denoted by
$A=\int_{-\infty}^{\infty} d\omega I(\omega)$. The power-Gaussian  
spectrum is similar to the original model spectrum studied by Caldeira and 
Leggett. This spectrum, along with the power exponential, is one of the most 
frequently used spectral models nowadays. The spectral area in 
(\ref{powGauss}) can be calculated analytically for different values of the  
Ohmicity parameter $\nu$ and the Gaussian cut-off $\Lambda$. The  
normalization with respect to the spectral area is included in the overall 
constant ${\cal A}$. 

The relaxation, dephasing and the leakage rates are calculated by 
examining $\rho_{22}^{(S)}(t), \vert\rho_{12}^{(S)}(t)\vert$ and 
$1-[\rho_{11}^{(S)}(t)+\rho_{22}^{(S)}(t)]$ respectively at   
short times $1 \le t \le 2.5$ [in units of the inverse absolute 
energy scale in ({\ref{hamilt.0})] where the dominant behavior is 
well-fitted to exponential. 

\subsection{The influence of the spectrum on RDL} 

The decoherence rates $(\tau_{_{RDL}})^{-1}$ corresponding to the  
qubit subspace of the system found by the numerical solution
of (\ref{densmatr.2}) are studied below. In Fig.(\ref{omega_0})
the rates are given for a varying spectral center $\omega_0$ with 
fixed spectral
areas. The figure suggests that the decoherence rates 
 in the qubit subspace are largely independent from
the center of the spectrum for the model spectra in
Eq's.(\ref{noisecorrelator.1}) and (\ref{noisecorrelator.2}) and
for the indicated range of spectral areas.
\begin{figure}
\includegraphics[scale=0.45,angle=-90]{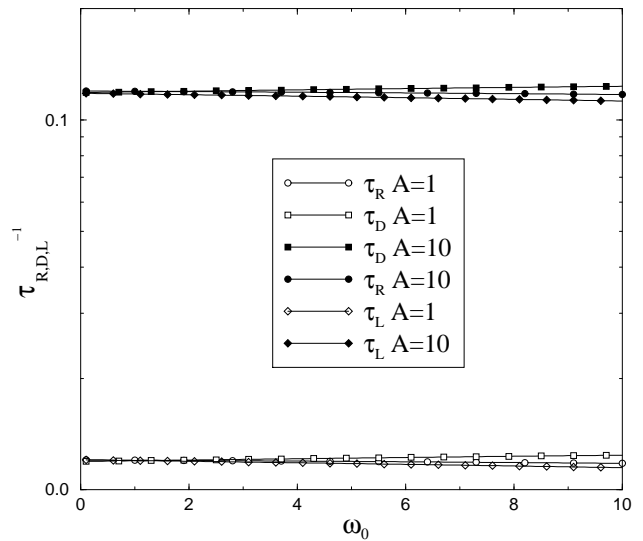}
\caption{Dependence of the RDL rates on the spectral shifts for
various spectral areas. Note that the vertical axis is
logarithmic.}
\label{omega_0}
\end{figure}
In Fig.(\ref{area}) and for the same three level system the RDL
rates are plotted against the spectral area for the rectangular
spectrum. The curves obtained for the Lorentzian spectrum totally
coincide with the corresponding ones for the rectangular spectrum
and they are not shown. On the other hand, the same quantities 
are also plotted in Fig.\,(\ref{powerGauss}) for the data obtained 
from the power-Gaussian spectrum for $\nu=-1,0,1$. The shown 
relaxation rates in 
Fig.\,(\ref{powerGauss}) are indistinguishable from those in 
Fig.\,(\ref{area}) within a very large range of spectral areas 
(The calculated dephasing and the leakage rates are in the same order of 
magnitude as the relaxation rates which can also be followed from 
Fig.\,(\ref{area}); hence, they are not shown in the figure).  
These observations prove the independence from the shape of the spectrum. 
Moreover, 
the plots (a) for $\nu=1$, (b) for $\nu=0$ and (c) for $\nu=-1$ in
Fig.(\ref{powerGauss}) yield identical decoherence rates for the same
horizontal scale. Hence, in the studied range $-1 \le \nu \le 1$ 
corresponding to  
sub-Ohmic, Ohmic and super-Ohmic (i.e. $\nu=-1,0,1$ respectively) 
the decoherence rates are determined solely by the spectral area. 
This observation is contrary to the earlier belief\cite{SB1} that 
it is the low frequency sector in the environmental spectrum 
[dominated by the $\omega^{1+\nu}$ term in (\ref{powGauss})] dominating 
the RDL rates.  

We hence observe independence from the shape and the
type of spectrum within a large range of spectral areas covering three  
orders of magnitude. The log-log axes in Fig's\,(\ref{area}) and 
(\ref{powerGauss})  
additionally indicate that the RDL rates depend linearly on the
spectral area. The monotonic dependence on the spectral area (whether 
it is a linear dependence or not), is a strong signature for the dominance  
of the non-resonant processes in the decoherence rates. 
\begin{figure}
\includegraphics[scale=0.5,angle=0]{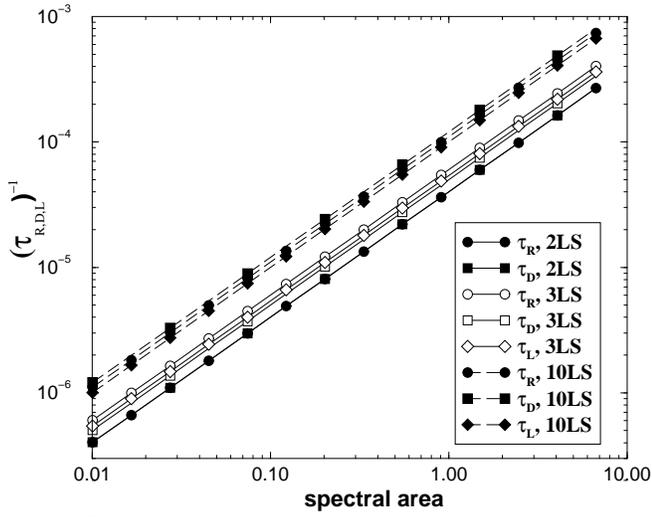}
\caption{Dependence of the RDL rates on the spectral area for the spectral 
function (\ref{noisecorrelator.1}).
Note that the both axes are logarithmic. In generating the ten level 
results we assumed that the system's energies above the first two levels 
are harmonic and given by $E_n=(n-1/2)~,3 \le n$ and all the couplings 
are $\varphi_{n\,r}=0.1$ for $n+r=odd$ and zero otherwise.}
\label{area}
\end{figure}
\begin{figure}
\includegraphics[scale=0.5,angle=0]{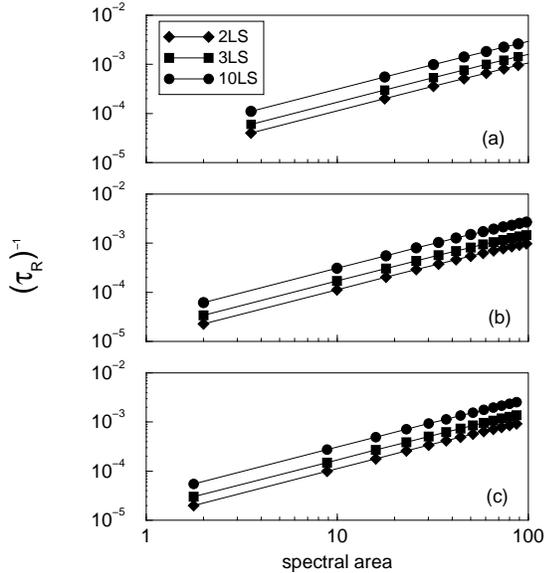}
\caption{Dependence of the relaxation rates on the spectral area for the 
spectral function (\ref{powGauss}) corresponding to  a) $\nu=1$ super-Ohmic, 
b) $\nu=0$ Ohmic, c) $\nu=-1$ sub-Ohmic spectra and for two, three and ten 
level systems. The trend for the dephasing and the leakage rates are 
very similar to Fig.(\ref{area}) and they are not shown here. 
Note that the both axes are logarithmic. The energies $E_n$ and couplings 
$\varphi_{n\,r}$ of the 
multilevel system are as indicated in Fig.(\ref{area}). The horizontal range 
in (c) is common to all graphs.}  
\label{powerGauss}
\end{figure}
Additionally, in Fig.(\ref{area}), and also implicitly in 
Fig.(\ref{powerGauss}), the leakage rates are shown to 
be in the same order of magnitude as the RD rates. The leakage rates 
increase by the inclusion of higher levels [also see Fig.(\ref{MLS_rates}) 
in Section II.B below]. Hence, we conclude that,  
the leakage demonstrates itself to be a non-negligible short time
effect manifested in multileveled systems.

The dominance of the non-resonant processes and their survival at
zero temperature implies that the influence of the higher levels   
 cannot be avoided independent
from how well the qubit is energetically separated from those 
levels. This observation is in contrast with the assumption that,
in low temperatures $T$ compared to the separation $\Delta E$ 
between the qubit subspace and higher states  
(i.e. $T \ll \Delta E$), the two level
approximation is well satisfied. 

\subsection{The influence of the multilevels on RDL}

The influence of the multilevels is examined by using a model 
coupling matrix $\varphi$ in (\ref{hamilt.0}) as 
\begin{equation}
\varphi_{n\,r}=0.1\,e^{-\vert n-r\vert/\Delta}~
\label{multicoupling}
\end{equation}
for $n+r=odd$ and zero otherwise. By varying the multilevel 
coupling range $\Delta$, the effect of the actively coupled levels on the  
RDL rates can be examined. In Fig.(\ref{MLS_rates}) the data is generated 
for different values of the multilevel range $\Delta=8, 40$ and the spectral 
area $A=5, 50$. The first observation is that, the linear trend observed in  
Fig.(\ref{area}) against the 
spectral area is respected within a larger range of multilevels. Secondly, 
the multilevel coupling range is observed to give rise to a saturation in   
the RDL rates at the onset $M \simeq \Delta$. Fig.(\ref{MLS_rates})  
therefore demonstrates further evidence of the observable effects 
in the decoherence rates arising from the finite coupling to multilevels.   

\begin{figure}
\includegraphics[scale=0.5]{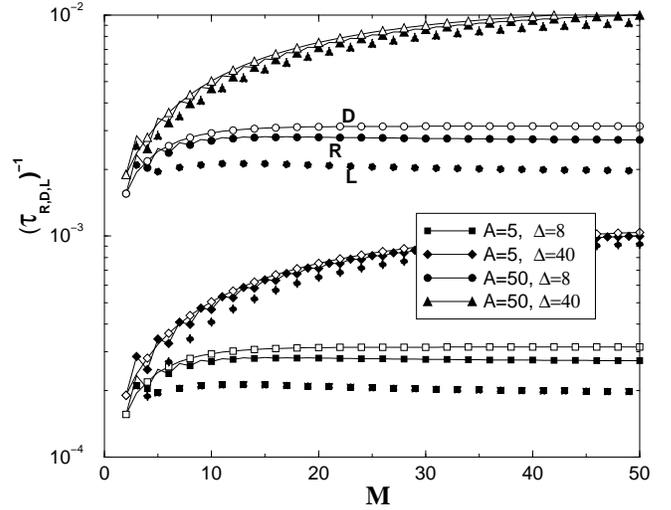}
\caption{RDL rates at $T=0$ against the number of
levels for different spectral areas and different multilevel 
coupling ranges $\Delta$. Here the coupling matrix elements 
between the levels $n$ and $r$ are modelled by 
$~\varphi_{n\,r}=0.1 ~ e^{-\vert n-r\vert/\Delta}~~$ for $n+r=odd$ an zero 
otherwise. Note the logarithmic vertical axis. The open and 
solid symbols superimposed on the solid line refer to dephasing and 
relaxation 
rates respectively. The pure symbols without connecting lines refer to 
leakage rates. Also $(A,\Delta)$ values are $(5,8)$ for squares, $(5,40)$ for 
diamonds, $(50,8)$ for circles and $(50,40)$ for triangles. Here, the 
Lorentzian spectrum is used.}  
\label{MLS_rates}
\end{figure}

\subsection{The fluctuations in the noise vacuum}

A characteristic feature of the coordinate coupling is to 
produce weak
vacuum fluctuations as well as finite number of environmental modes 
at zero temperature. These corrections,  
 can be calculated perturbatively and they give a clue
about the effectiveness of the virtual (non-resonant) processes. 

The corrections to environmental number of modes can be 
calculated using the retarded Greens function  
\begin{equation}
G_\omega(t-t')=-i \Theta(t-t^\prime)\, \langle {\cal T} 
\tilde{a}_\omega^{\dag}(t)\tilde{a}_\omega(t^\prime)\tilde{\cal
S}(\infty,-\infty)\rangle
\label{greens}
\end{equation}
where ${\cal T}$ denotes the time ordering, the $~\tilde{}~$ denotes
the interaction picture and the ${\cal S}$ matrix includes the
interaction Hamiltonian in (\ref{hamilt.0}) in the interaction
picture. The environmental {\it photon number} is than found by
the standard method $n_{\omega}=i\lim_{t^\prime \to t}\,G_\omega(t-t')$.
The average is taken over the noninteracting initial
state including system and the environment in the product form
\begin{equation}
|\rangle=(a|1\rangle+b|2\rangle)\otimes|env\rangle.
\end{equation}
where $\vert env\rangle$ is characterized by the spectral function 
$I(\omega)$ at zero temperature. Evaluating the time ordered  
 integrals arising from the expansion of the ${\cal S}$ matrix,
and taking the limit $t'\rightarrow t$ we obtain a second order
correction as
\begin{equation}
\begin{array}{lrlr}
{n}_\omega^{(2)}&&=\sum_{s=1}^{3} \frac{2 I(\omega)}{\lambda_s^2} \\
&&\times (|a|^2\varphi_{0s}+|b|^2\varphi_{1s}+
(a^*b+b^*a)\varphi_{1s}\varphi_{0s})
\label{n_omega.1}
\end{array}
\end{equation}
where $\lambda_s=\omega+E_1-E_s$ and and $E_1=E_2$ as announced
in (\ref{hamilt.0}). Eq.(\ref{n_omega.1}) yields a divergent 
result for the total number of photons even for arbitrarily small 
system-environment couplings. The divergence 
is obtained independently from the type of the spectrum in
(\ref{noisecorrelator.1}) or (\ref{noisecorrelator.2}) and arises
from the second order singularity in (\ref{n_omega.1}). To avoid
this unphysical
 result in the second order, we convert the second order correction
into an RPA sum following conventional practices in perturbative
approaches\cite{Mahan}. As we use RPA, the second order pole in
(\ref{n_omega.1}) is split into two first order poles which yields
a finite result at the RPA level. The relevant RPA graphs are 
depicted in Fig.\,(\ref{RPA}) below. 
\begin{figure}
\includegraphics[scale=0.4,angle=0]{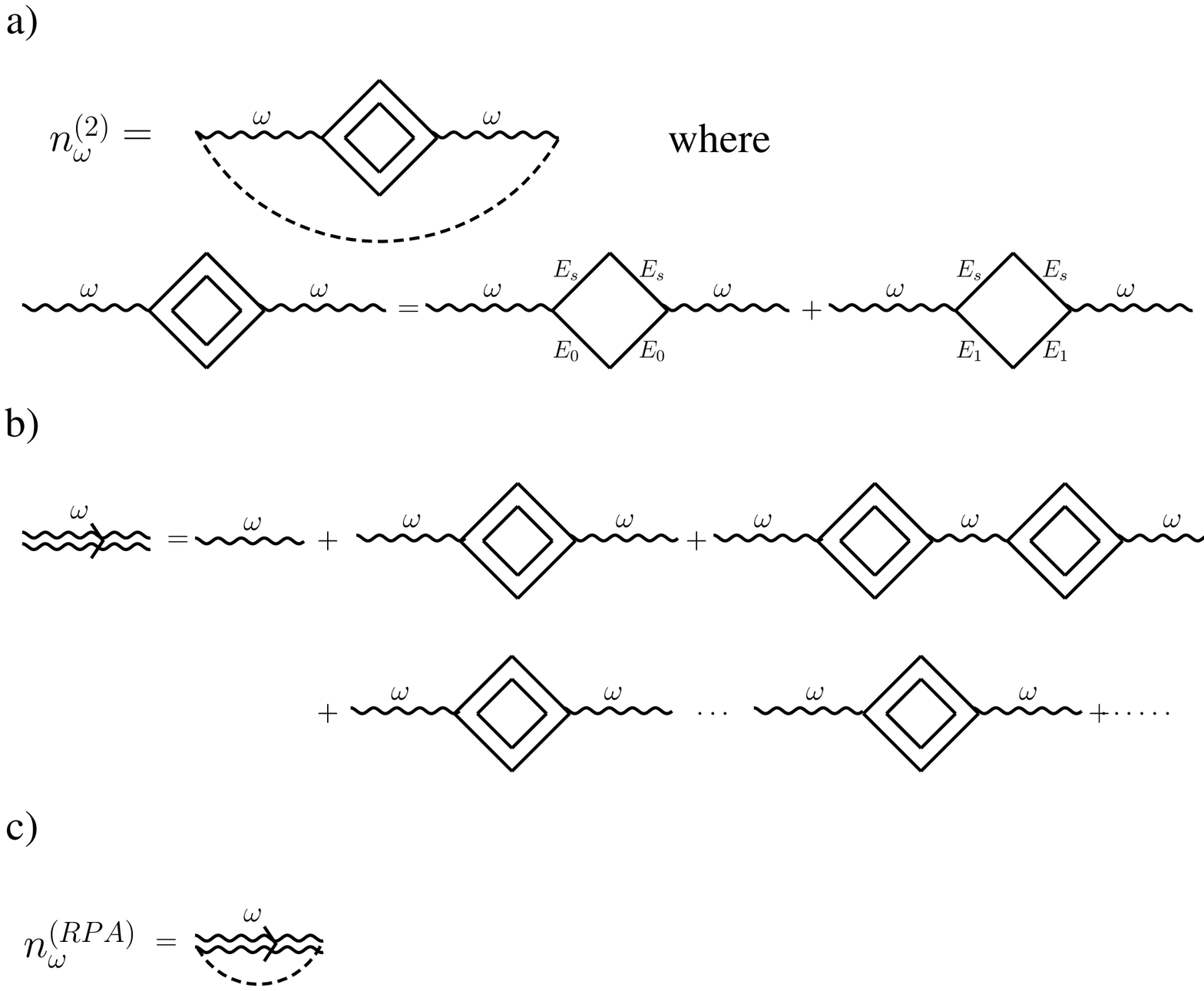}
\caption[cap419]{Contributing diagrams in the calculation of
(\ref{greens2}). (a)
The definition of the basic second order root diagrams that are used
to implement RPA. The dashed arc lines indicate the retarded limit
$t^\prime \to t+0^{-}$ in the calculation of $n_\omega^{(2)}$,
(b) the RPA hierarchy using the root diagram in (a),
 (c) the resulting RPA Green's function used in the calculation of
(\ref{fluctuation}). Here the dashed arc lines indicate again the
retarded limit $t^\prime \to t+0^{-}$ in the calculation of
$n_\omega^{(RPA)}$.}
\label{RPA}
\end{figure}

Next, we calculate the fluctuations in the number of photons
$\Delta n_\omega$ by using a four point retarded Green's function
for the environment as
\begin{equation}
D_\omega(t-t')=-i \Theta(t-t^\prime)\, \langle {\cal T} 
\tilde{a}_k^{\dag}(t)\tilde{a}_\omega(t)
\tilde{a}_\omega^{\dag}(t^\prime)\tilde{a}_k(t^\prime)
\tilde{\cal S}(\infty,-\infty)\rangle
\label{greens2}
\end{equation}
The connected part of
(\ref{greens2}) directly yields the fluctuations in the photon number 
as $\Delta n^2_\omega=i \lim_{t^\prime \to t}\,D_\omega(t-t')$. The
diagrams corresponding to the RPA scheme using (\ref{greens2}) are shown  
in Fig.(\ref{RPA}). The fluctuation in the total photon number is  
found by $(\Delta N)^2=\int d\omega\,\Delta n^2_\omega$
and is given by
\begin{equation}
(\Delta N)^2=\sum_s
\int_{-\infty}^{\infty}\frac{d\omega}{(\omega+\lambda_s)}
\frac{1}{1-4(system)_sI(\omega)/(\omega+\lambda_s)}
\label{fluctuation}
\end{equation}
which is plotted in
Fig.(\ref{fluct}) with respect to $\epsilon$ and parameterized by
$\omega_0$ for the two, three and ten level systems. The bare spectral
area is fixed at $A=5\pi$. The observed
independence from $\epsilon$ at fixed spectral area implies
independence from the spectral shape. Additionally independence
from $\omega_0$ is also shown. 

\begin{figure}
\includegraphics[scale=0.4,angle=-90]{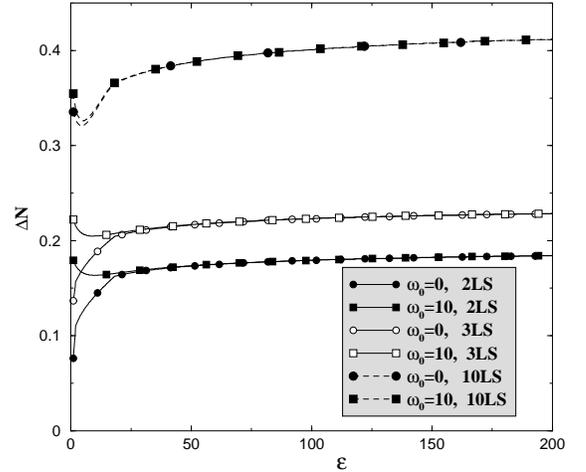}
\caption[cap419]{Fluctuations in the total photon number versus
the spectral width for various $\omega_0$ values computed for two
three and ten leveled MLS. Here the bare spectral area $A=5\pi$}
\label{fluct}
\end{figure}

In Fig.(\ref{DeltaN_N.vs.area}) below, the ratio $\Delta N/N$ is plotted 
as a function of the spectral area. The figure, together with the inlet 
for $\Delta N$ versus spectral area, is well fitted to  
\begin{equation}
\frac{\Delta N}{N} \simeq \frac{Const.}{\sqrt{A}}
\label{DeltaN_N.vs_area}
\end{equation}
We observe that $\Delta N \propto \sqrt{A}$ and $N \propto A$ in a large 
range of spectral areas. Here the positive constant $Const.$ depends on the 
number of multi levels.   
\begin{figure}
\includegraphics[scale=0.4,angle=0]{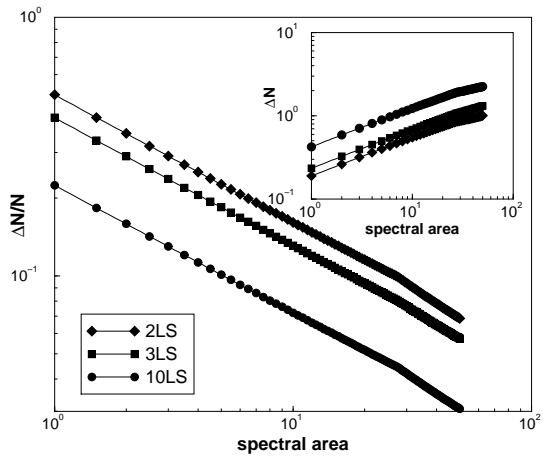}
\caption[cap419]{The ratio $\Delta N/N$ versus the spectral area 
(inlet: $\Delta N$ versus the spectral area).}  
\label{DeltaN_N.vs.area}
\end{figure}

The results of this subsection are in a sense also a check 
for the validity of the Born-Oppenheimer approximation. The 
correction obtained to the spectral area as a result of the 
system-environment coupling is bounded by a few ten 
percent. However, as the number of levels increase the total fluctuations  
 decrease relative to the increase in the total number of environmental modes. 
For a MLS with strong couplings between a large 
number of levels the validity of the Born-Oppenheimer approximation may  
still survive although we did not check this result explicitly. 

\section{Conclusions}
It is shown that for the most conventionally used system-environment
couplings of linear coordinate type the non-resonant processes overwhelm 
the resonant ones in their contribution to the decoherence rates at 
zero temperature. In this regard, and within the studied spectral range, 
it is observed that the decoherence rates do not depend on the  
specific low/high energy properties of the spectrum, whereas, a strong 
(and dominantly) dependence is observed on the overall spectral area. These 
results are confirmed for three independent spectral profiles.   

We have also examined the effect of the system-environment coupling.  
We observed that the number of levels which are coupled by the 
system-environment coupling plays a non-negligible role
in decoherence time rates. We see
that as the number of  coupled levels in the system increase, both
the decoherence time rates and photon number corrections increase
further from the rates and corrections in a pure two level system.
The observed effects for the MLS cannot be explained by an equivalent
renormalization of the two levelled subsystem. In the light of these 
observations, we conclude that the conventional postulates  
 of the two-level approximation do not necessarily lead 
to a dynamical behaviour of a multilevelled system largely confined to 
its lowest two levels. The presented results are obtained from the 
numerical solution of the master equation with the Born-Oppenheimer 
approximation, and hence, they are exact in the short time limit where 
the decoherence properties are studied.   
\newpage
\section{Acknowledgements}
This research is supported by the Scientific and the Technical
Research Council of Turkey (T\"{U}B{\.I}TAK) grant number
TBAG-2111 (101T136) and by the Dicle University grant number
D\"{U}APK-02-FF-27. E. Me\d{s}e thanks the Bilkent University for
hospitality. We thank C. Sevik for his help with the Fig.(\ref{RPA}).

\end{document}